\documentclass[12pt, a4paper]{article}

\usepackage[T1]{fontenc}
\usepackage[utf8]{inputenc}
\usepackage[english]{babel}

\usepackage[top=2cm, bottom=2cm, left=2cm, right=2cm]{geometry}
\usepackage{fancyhdr}
\usepackage{setspace}
\usepackage{amsfonts,amssymb,amsthm,amsmath,amscd,mathtools}
\usepackage{dsfont}
\usepackage{bm}
\usepackage{esint}
\usepackage{enumitem}

\usepackage{subcaption}

\usepackage[table]{xcolor}
\usepackage{graphicx}
\usepackage{tikz}
\usetikzlibrary{positioning, shapes.geometric, arrows.meta}

\usepackage{natbib}
\usepackage{newtxtext}
\usepackage[subscriptcorrection]{newtxmath}
\usepackage[plain,noend]{algorithm2e}
\usepackage[normalem]{ulem}
\usepackage{comment}

\usepackage[citecolor=blue]{hyperref}

\theoremstyle{definition}
\newtheorem{theorem}{Theorem}[section]

\newtheorem{assumption}[theorem]{Assumption}

\newcommand{\T}{\mathrm{\scriptscriptstyle T}}

\newcommand{\pr}{\operatorname{pr}}

\newcommand{\tr}{\operatorname{tr}}

\newcommand{\FDP}{\textsc{fdp}}
\newcommand{\TDP}{\textsc{tdp}}

\newenvironment{keywords}
{\noindent\textbf{Keywords: }}
{\par}

\usepackage{titlesec}
\titleformat{\section}
{\normalfont\Large\bfseries}
{\thesection}{1em}{}
\titleformat{\subsection}
{\normalfont\large\bfseries}
{\thesubsection}{1em}{}
\titleformat{\subsubsection}
{\normalfont\normalsize\bfseries}
{\thesubsubsection}{1em}{}

\title{\textbf{A Bayesian framework with adaptive elastic nets for the inference of Gaussian graphical models}}
\author{
	\textbf{Roland B. Sogan}\\
	\small Sorbonne Université, Université Paris Cité, CNRS,\\
	\small Laboratoire de Probabilités, Statistique et Modélisation,\\
	\small \href{mailto:roland-boniface.sogan@sorbonne-universite.fr}{\texttt{roland-boniface.sogan@sorbonne-universite.fr}}
	\and
	\textbf{Tabea Rebafka}\\
	\small MIA AgroParisTech, INRAE, Université Paris-Saclay,\\
	\small \href{mailto:tabea.rebafka@agroparistech.fr}{\texttt{tabea.rebafka@agroparistech.fr}}
	\and
	\textbf{Fanny Villers}\\
	\small Sorbonne Université, Université Paris Cité, CNRS,\\
	\small	Laboratoire de Probabilités, Statistique et Modélisation,\\
	\small \href{mailto:fanny.villers@upmc.fr}{\texttt{fanny.villers@upmc.fr}}
}
\date{}
\begin{document}
	
	\maketitle
	
	
\begin{abstract}
	Estimating conditional independence graphs from high-dimensional Gaussian data is challenging because methods must detect relevant edges while rigorously controlling statistical errors. We propose a Bayesian framework based on a prior accounts for degree heterogeneity edge sparsity, and graph topology the graph. The resulting posterior distribution is incorporated into a multiple testing procedure for graph inference with false discovery rate control. Computation is carried out through a combination of adaptive elastic nets and a variational expectation--maximization algorithm. In simulations, the method achieves reliable false discovery rate control while maintaining strong power, especially in heterogeneous networks such as graphs with hubs, and remains competitive under structural misspecification. Applications to breast cancer gene expression data and financial return networks show that the method yields sparse and interpretable conditional dependence graphs while retaining the most stable interactions detected by competing approaches.
	
\end{abstract}

	\section{Introduction}
	
	\subsection{Problem statement}

	Understanding dependence structure among variables is a central problem in modern statistical analysis. In many scientific applications, the aim is not merely to quantify marginal association, but to identify the pattern of direct interactions governing the joint behaviour of a multivariate system. Graphical representations provide a natural and interpretable summary of such structure, with variables represented by nodes and relationships by edges. Gaussian graphical models \citep{Lauritzen1996, Koller2009} provide a standard framework for learning such structure from multivariate Gaussian data. Owing to their interpretability and flexibility, they have been widely used in genomics, ecology, neuroscience and finance \citep{Shutta2022, Ranciati2021, DaFermo2024, Diebold2014, Smith2011, Varoquaux2010}.
	
	In high-dimensional settings, where the number of variables exceeds the sample size, substantial work has been devoted to inference in Gaussian graphical models; see, for example, \citet{Meinshausen2006}, \citet{Friedman2008}, \citet{Liu2013}, \citet{Gan2019} and \citet{Kilian2024}. Despite these advances, it remains difficult to control statistical errors while detecting as many true edges as possible. This issue is important in practice, because false discoveries may lead to incorrect scientific conclusions and costly empirical validation. In this paper we consider control of the false discovery rate (FDR) \citep{Benjamini1995}, defined as the expected proportion of false discoveries among selected edges. Our aim is to develop a method for edge detection that controls the FDR while retaining high power.
	
	Our approach is developed in a Bayesian framework, in which estimation yields posterior probabilities of edge inclusion. Such quantities have already been used for graph recovery, for example by \citet{Gan2019}, who proposed the BAGUS method and recovered the graph by thresholding posterior probabilities at $0.5$. Building on ideas from multiple testing, we use posterior edge probabilities to construct a procedure that controls the FDR at a user-specified nominal level. To improve power, we account for structural heterogeneity in graph topology, motivated by the fact that networks arising in applications often exhibit hubs, community structure and varying connectivity patterns.
	
	\subsection{Related work}
	\label{sec:related-works}
	
	In Gaussian graphical models, the primary object of interest is conditional independence as marginal correlations may reflect indirect associations induced by other variables. Two variables are conditionally independent if and only if the corresponding entry of the precision matrix, defined as the inverse covariance matrix, is zero. The conditional independence graph is therefore determined by the non-zero off-diagonal entries of the precision matrix.
	
	A major class of methods for inferring conditional independence graphs in high-dimensional Gaussian graphical models is based on penalized likelihood estimation of the precision matrix $K$, where the penalty induces sparsity. These methods can be written in the form
	\[
	\widehat K = \arg\max_{K \succ 0} \bigl\{ \ell(K;S) - \mathrm{pen}(K)\bigr\},
	\]
	where $\ell(K;S) = n(\log\det K - \operatorname{tr}(SK))/2$ up to an additive constant, the Gaussian loglikelihood based on the sample covariance matrix $S$, and $\mathrm{pen}(K)$ is a penalty function that encourages sparsity in the graph structure. A prominent example is the graphical lasso \citep{Yuan2007, Banerjee2008, Friedman2008}, which uses an element-wise $\ell_1$ penalty,
	\[
	\mathrm{pen}(K) = \lambda \|K\|_1,\quad
	\|K\|_1 = \sum_{i<j} |K_{ij}|.
	\]
	This penalty shrinks some coefficients to zero and leads to computationally efficient procedures, but it treats all entries of $K$ equally, regardless of their structural role in the network. Such uniform penalization may be suboptimal when the underlying graph is heterogeneous.

	To address this issue, \citet{Ambroise2008} proposed SIMoNe, which introduces latent groups labels $Z_i \in \{1, \ldots, Q\}$ for each node $i$ and uses a block-weighted $\ell_1$ penalty of the form $\mathrm{pen}(K) = \|\lambda_Z(K)\|_1$. Here, $\lambda_Z(K)$ denotes a group-dependent weighting of the entries of $K$, so that the amount of penalization applied to $K_{ij}$ depends on the groups to which nodes $i$ and $j$ belong, thereby allowing different degrees of penalization within and between groups. This encourages heterogeneous sparsity patterns adapted to block structure. However, the method does not provide posterior probabilities for edges and so does not yield a mechanism for false discovery rate control.
	
	An alternative is to work in a Bayesian framework, where sparsity is induced through a prior distribution on the entries of $K$ rather than through an explicit penalty. \citet{Wang2012} showed that the graphical lasso estimator is the maximum a posteriori estimator under independent Laplace priors on the off-diagonal entries of $K$. The BAGUS method of \citet{Gan2019} is based on  a hierarchical Bayesian model in which the distribution of each entry $K_{ij}$ depends on whether the corresponding edge is present. It relies on a spike-and-slab prior formed by two Laplace distributions with different scales. The corresponding maximum a posteriori estimator can be viewed as arising from a non-convex penalty, $\mathrm{pen}(K) = \sum_{i<j} \mathrm{pen}_{\mathrm{SS}}(K_{ij})$, where $\mathrm{pen}_{\mathrm{SS}}(K_{ij}) = -\log \pi(K_{ij})$ and $\pi(K_{ij})$ denotes the spike-and-slab prior. Computation is made tractable through an expectation-maximization algorithm that solves a sequence of weighted graphical lasso problems. Unlike many competing approaches, BAGUS separates estimation of the precision matrix from inference on the graph: $K$ is first estimated, and the graph is then recovered by thresholding approximate posterior edge probabilities at $1/2$.
	
	For the purposes of this paper, BAGUS has two limitations. First, unlike SIMoNe, it does not exploit heterogeneity in graph topology. Second, thresholding posterior probabilities at $1/2$ provides no guarantee on the false discovery rate.
	
	Only a limited number of procedures explicitly control the false discovery rate in Gaussian graphical model selection. One of the earliest is the graphical false discovery control procedure of \citet{Liu2013}, based on multiple testing of conditional correlations. More recently, knockoff-based methods have been extended to Gaussian graphical models \citep{Li2021}, following the general framework of \citet{Barber2015}. These approaches provide finite-sample false discovery rate guarantees by constructing synthetic variables that preserve the dependence structure while enabling valid edge selection. Another recent contribution is due to \citet{Kilian2024}, who adapted the noisy stochastic block model framework of \citet{Rebafka2022} to the Gaussian graphical model setting and used graph-based test statistics to control the false discovery rate. These approaches also have limitations. Testing procedures such as that of \citet{Liu2013} may suffer from low power in high dimensions. Knockoff methods, although attractive theoretically, require the construction of valid knockoffs and are mainly developed for moderate-dimensional settings. The NSBM-based approach depends strongly on the quality of the input statistics, and its performance may deteriorate under model misspecification.

	\subsection{Contributions}
	
	We propose a Bayesian procedure for inferring conditional independence graphs in heterogeneous high-dimensional Gaussian graphical models under false discovery rate control. In the sequel, we refer to the proposed procedure as HSS-GGM, short for heterogeneous spike-and-slab Gaussian graphical model. Our main contributions are as follows.
	\begin{enumerate}
		\item We introduce a Bayesian model for conditional independence graph estimation that combines a stochastic block model for heterogeneous graph structure with a spike-and-slab prior on the precision matrix. This yields an edge-specific adaptive elastic-net procedure, in which the amount of shrinkage depends on posterior edge probabilities.
		
		\item We develop a multiple testing procedure for graph selection that uses posterior edge probabilities to control the false discovery rate.
		
		\item We develop a computationally tractable inference algorithm that alternates between variational estimation of the latent block structure and model parameters, and adaptive elastic-net estimation of the precision matrix.
		
		\item Numerical experiments on simulated data show that the proposed method controls the false discovery rate at nominal levels across different graph topologies while maintaining high power and remaining robust to model misspecification. Applications to breast cancer gene expression data and financial return networks show that the method yields sparse and interpretable conditional independence graphs while retaining stable interactions detected by competing methods.
	\end{enumerate}

	\section{Bayesian formulation for Gaussian graphical models}
	\label{sec:bayesian-ggm-elasticnet}
	
	\subsection{Statistical model and notation}
	
	We consider the problem of recovering the conditional independence graph of a Gaussian random vector
	\[
	X=(X_1,\ldots,X_p)^\T \in \mathbb R^p,
	\qquad
	X\sim \mathcal N(0,\Sigma).
	\]
	Given \(n\) independent copies \(X^{(1)},\ldots,X^{(n)}\), we write
	\[
	\mathbf X=(X_{i,j})_{1\le i\le n,\ 1\le j\le p}\in\mathbb R^{n\times p}
	\]
	for the data matrix, whose \(i\)th row is \((X^{(i)})^\T\). Hence \(X_j\) is the \(j\)th component of \(X\), \(X_j^{(i)}\) is the \(j\)th component of the \(i\)th observation, and \(X_{i,j}=X_j^{(i)}\). Let \(K = \Sigma^{-1}\) denote the corresponding precision matrix. In a Gaussian graphical model, conditional independence is equivalent to a zero off-diagonal entry of \(K\), that is,
	\begin{equation}
		X_i \perp\!\!\!\perp X_j \mid \mathbf X_{-\{i,j\}}
		\quad \Longleftrightarrow \quad
		K_{ij} = 0.
	\end{equation}
	We denote the support of \(K\) by the binary matrix \(A = (A_{ij})\), defined for \(i \neq j\) by
	\[
	A_{ij} = \mathds{1}_{\{K_{ij} \neq 0\}}, \qquad A_{ii} = 0.
	\]
	Thus \(A\) is the adjacency matrix of the conditional independence graph associated with \(K\) where nodes correspond to variables and edges encode conditional dependencies between them. Recovering this graph may be viewed as a multiple testing problem, in which each potential edge corresponds to the hypotheses
	\begin{equation}
		\label{eq:hypothesis}
		H_{0,ij}: A_{ij} = 0
		\qquad \text{versus} \qquad
		H_{1,ij}: A_{ij} = 1,
	\end{equation}
	for all \((i,j)\in\mathcal A\), where
	\(
	\mathcal A=\{(i,j):1\le i<j\le p\}
	\).
	A Bayesian hierarchical model with prior distributions on the graph \(A\) and the precision matrix \(K\) provides a natural framework for computing posterior probabilities of edge inclusion and for deriving multiple testing rules from these posterior quantities.

	\subsection{Hierarchical spike-and-slab prior on the precision matrix}
	
	We adopt a Bayesian approach to test the hypotheses in \eqref{eq:hypothesis}, using a prior that allows larger values of \(K_{ij}\) when an edge is present and stronger concentration around zero when no edge is present. This leads naturally to a spike-and-slab formulation.
	
	Spike-and-slab priors are a Bayesian tool for inducing sparsity. They were introduced by \citet{Mitchell1988} as a mixture of a point mass at zero, called the spike, and a diffuse distribution for non-zero coefficients, called the slab. Although conceptually appealing, such discrete formulations generally lead to difficult posterior computation. Continuous spike-and-slab priors address this issue by replacing the Dirac spike with a sharply concentrated continuous distribution. A notable example is the spike-and-slab lasso of \citet{Rovckova2018,Rovckova2018spike,Rovckova2016}, which uses two Laplace components with different scales and yields tractable penalized likelihood estimators in sequence and regression models. In the context of Gaussian graphical models, \citet{Gan2019} adopted  a spike-and-slab prior formed by two Laplace distributions  for  estimation of the precision matrix. While such a formulation preserves computational tractability, a Laplace slab may induce substantial shrinkage of large signals.
	
	To alleviate this effect, we replace the Laplace slab with a Gaussian slab. The Laplace spike preserves strong concentration around zero, whereas the Gaussian slab has a softer shrinkage effect on large coefficients and is therefore better suited to retaining strong edge signals. More precisely, for each off-diagonal entry \(K_{ij}\), \(i < j\), we define
	\begin{equation}
		K_{ij} \mid A_{ij} \sim
		\begin{cases}
			\mathrm{Laplace}(\xi_0), & \text{if } A_{ij} = 0 \quad (\text{spike}),\\
			\mathcal{N}(0, \sigma_1^2), & \text{if } A_{ij} = 1 \quad (\text{slab}),
		\end{cases}
		\label{eq:prior_K_given_A}
	\end{equation}
	where the Laplace distribution with rate parameter \(\xi_0 > 0\) has density
	\[
	g_{\xi_0}(x) = \frac{\xi_0}{2}\exp(-\xi_0 |x|), \qquad x \in \mathbb{R},
	\]
	and the Gaussian density with mean \(0\) and variance \(\sigma_1^2\) is
	\[
	\phi(x;0,\sigma_1^2)=\frac{1}{\sqrt{2\pi}\,\sigma_1}\exp\!\left(-\frac{x^2}{2\sigma_1^2}\right),
	\qquad x\in\mathbb R.
	\]
	For the diagonal entries, which do not correspond to edges, we follow \citet{Deshpande2017} and assign independent exponential priors
	\begin{equation}
		K_{ii} \sim \mathrm{Exponential}(\sigma_1),
		\qquad i = 1, \ldots, p.
		\label{eq:prior_diag}
	\end{equation}
	The quantities $\xi_0$ and $\sigma_1$ are hyperparmeters whose calibration is discussed in Section \ref{}.
	We also impose the positive-definiteness constraint \(K \succ 0\). 
	This completes the specification of the conditional prior on \(K \mid A\). It  remains to specify a prior distribution for the graph \(A\).
	
	\subsection{Structured prior on the conditional independence graph}
	
	A simple choice is to draw the edge indicators \(A_{ij}\) independently from a \(\mathrm{Bernoulli}(\eta)\) distribution with common success probability \(\eta\), as is assumed in many sparse graphical model priors \citep{Gan2019,Vogels2024,Xi2024}. However, such a homogeneous prior does not capture the heterogeneous connectivity patterns often observed in real networks, where some nodes act as hubs, others remain peripheral, and connection probabilities vary across the graph.
	
	To account for this heterogeneity, we model the graph \(A\) through a stochastic block model \citep{Holland1983,Nowicki2001}. Each node is assigned to one of \(Q\) latent groups according to
	\begin{equation}
		Z_i \sim \mathrm{Multinomial}(1, \pi),
		\qquad i = 1, \ldots, p,
		\label{eq:prior_Z}
	\end{equation}
	where \(\pi = (\pi_1, \ldots, \pi_Q)\) satisfies \(\pi_q > 0\) and \(\sum_{q=1}^Q \pi_q = 1\). Conditionally on the labels \(Z=(Z_1,...,Z_p)\), the edge indicators are independent and satisfy
	\begin{equation}
		A_{ij} \mid (Z_i = q, Z_j = \ell) \sim \mathrm{Bernoulli}(\omega_{q\ell}),
		\qquad (i,j) \in \mathcal{A},
		\label{eq:prior_A_given_Z}
	\end{equation}
	where \(\omega \in (0,1)^{Q \times Q}\) is symmetric and \(\omega_{q\ell}\) denotes the connection probability between groups \(q\) and \(\ell\).
	We denote by $\theta = (\pi, w) $ the unknown model parameters arising from the stochastic block model.

	Combining \eqref{eq:prior_K_given_A}--\eqref{eq:prior_A_given_Z} gives the joint hierarchical prior
	\begin{equation}
		\Pi(K, A, Z)
		=
		\left[\prod_{i=1}^{p} \Pi(Z_i)\right]
		\left[\prod_{i<j} \Pi(A_{ij} \mid Z_i, Z_j)\right]
		\left[\prod_{i<j} \Pi(K_{ij} \mid A_{ij})\right]
		\left[\prod_{i=1}^{p} \Pi(K_{ii})\right]
		\mathds{1}_{\{K \succ 0\}}.
		\label{eq:joint_prior_KAZ}
	\end{equation}
	

\section{Posterior approximation and inference}

\subsection{Inference strategy by alternating optimization}

Our primary inferential goal is graph recovery, that is, deciding whether an
edge is present between two nodes \(i\) and \(j\). Under the proposed
hierarchical model, the natural Bayesian target is the posterior edge
probabilities
\begin{equation*}
	\pr(A_{ij}=1 \mid \mathbf X),
	\qquad 1 \le i < j \le p,
	\label{eq:posterior_edge_prob}
\end{equation*}
which are obtained by integrating out the latent precision
entries $K_{i,j}$ and the latent block memberships $Z_{ij}$
\begin{equation*}
	\pr(A_{ij}=1 \mid \mathbf X)
	=
	\sum_{q,\ell}
	\int
	\pr(A_{ij}=1 \mid K_{ij}, Z_i=q, Z_j=\ell)\,
	\Pi(K_{ij}, Z_i=q, Z_j=\ell \mid \mathbf X)\, dK_{ij}.
	\label{eq:posterior_edge_integral}
\end{equation*}
Obviously, these quantities are difficult to estimate. Instead, our procedure aims to estimate the conditional
posterior probabilities
\[
\pr(A_{ij}=0 \mid K_{ij}, Z_i, Z_j; \theta), 
\]
which admit a closed-form expression and can be evaluated once \(K\), the
latent blocks $Z$ and the model parameters $\theta$ have been estimated. Note that in the hierarchical model, the latent graph $A$ and the observations $\mathbf X$ are conditionally independent given $K$, which implies that the  posterior distribution of $A$ given $(\mathbf{X}, K)$ does not depend on $\mathbf{X}$, that is  \(\Pi(A\mid \mathbf X,K)=\Pi(A\mid K)\).
To jointly estimate all unknown quantities in the model,  we propose an alternating optimization algorithm in which each set of unknown quantities is updated while keeping the others fixed. 
The algorithm consists of two steps. 
In the first step, when \(K\) is fixed, the model corresponds to a so-called noisy stochastic block model, and a variational expectation--maximization (VEM) algorithm can be used to fit the model to the data \citep{Rebafka2022}. This step relies on a variational approximation of the conditional distribution of \((A,Z)\) given \(K\). The algorithm provides estimates of the latent block memberships \(Z\) in addition to parameter estimates. The second step updates the precision matrix \(K\) by maximizing a penalized likelihood criterion, where the penalty takes the form of a weighted elastic net induced by our spike-and-slab prior. The \(\ell_1\) component, induced by the Laplace spike, promotes sparsity by shrinking coefficients toward zero, while the \(\ell_2\) component, induced by the Gaussian slab, allows for larger values. The penalty is edge-specific, with weights given by the posterior edge probabilities, resulting in adaptive shrinkage across edges. The two steps of the inference algorithm are described in more detail in the following two subsections.

\subsection{Variational expectation--maximization}\label{section::VEM}

The general VEM algorithm for inference on \((A,Z)\) and the parameters \(\theta\), for a fixed precision matrix \(K\), was first proposed in \cite{Rebafka2022}. It is adapted here to the specific choice of Gaussian and Laplace distributions for the precision matrix entries. Globally, the VEM algorithm maximizes a lower bound on the marginal log-likelihood \(\log \Pi(\mathbf X,K;\theta)\), known as the evidence lower bound,
\[
\mathcal{J}(\mathbf X,K, \mathcal Q;\theta) = \mathbb{E}_{\mathcal{Q}}\bigl[\log \Pi(\mathbf X,K,A,Z;\theta)\bigr] + \mathcal{H}(\mathcal{Q}),
\]
where \(\mathcal{Q}(A,Z)\) denotes a variational distribution approximating the conditional distribution of the latent variables \((A,Z)\) given \(K\) under the current value of \(\theta\), and \(\mathcal{H}(\mathcal{Q})\) denotes its entropy. The algorithm alternates between two steps until convergence. The M-step maximizes \(\mathcal{J}\) with respect to \(\theta\) while keeping \(\mathcal{Q}\) fixed. The E-step maximizes \(\mathcal{J}\) with respect to \(\mathcal{Q}\) while keeping \(\theta\) fixed, which is equivalent to minimizing the Kullback--Leibler divergence
\[
\mathrm{KL}(\mathcal{Q}(A,Z)\,\|\,\Pi(A,Z\mid X,K;\theta)),
\]
thereby yielding the best approximation to the conditional distribution of the latent variables. A common choice for the variational distribution is the mean-field approximation \citep{Jordan1999,Daudin2008,Blei2017}, using the class of factorized distributions of the form
\begin{equation}
	\mathcal{Q}(A,Z)
	=
	\Pi(A\mid K,Z;\theta)\prod_{i=1}^p \tau_{i,Z_i},
	\label{eq:variational_family}
\end{equation}
where \(\tau_{iq}\in[0,1]\) and \(\sum_{q=1}^Q \tau_{iq}=1\). The parameter \(\tau_{iq}\) represents the posterior probability that node \(i\) belongs to block \(q\). The variational E-step provides updates of all variational parameters \(\tau=(\tau_{iq})_{1 \le i\le p,\ 1\le q\le Q}\), by solving a fixed-point equation, as well as estimates of the posterior edge probabilities
\begin{equation}
	\rho_{q\ell}^{\,i,j}
	:=
	\pr(A_{ij}=1\mid K_{ij},Z_i=q,Z_j=\ell; \theta)
	=
	\frac{\omega_{q\ell}\phi(K_{ij};0,\sigma_1^2)}
	{\omega_{q\ell}\phi(K_{ij};0,\sigma_1^2)
		+
		(1-\omega_{q\ell})g_{\xi_0}(K_{ij})},
	\label{eq:rho_update}
\end{equation}
for each node pair \((i,j)\) and block pair \((q,\ell)\), with \(\phi(\cdot;0,\sigma_1^2)\) denoting the Gaussian density with mean \(0\) and variance \(\sigma_1^2\), and \(g_{\xi_0}(\cdot)\) denoting the Laplace density with rate parameter \(\xi_0\). More details are provided in Appendix~\ref{app:vem-derivations}.

\subsection{Precision matrix estimation by adaptive elastic-net regression}\label{section::updateK}
The precision matrix $K$ is updated by maximizing the expected complete-data 
log-likelihood under the current variational distribution
\begin{equation}
	\mathbb{E}_{\mathcal{Q}}\bigl[ \log \Pi(\mathbf X,K,A,Z;\theta) \bigr].
	\label{expectedcomplete}	
\end{equation}
This expectation effectively approximates the unknown latent variables using their estimated posterior distributions. 
Maximizing \eqref{expectedcomplete} with respect to $K$ amounts to maximizing
\begin{equation}
	\label{eq:evidence}
	\ell(K;S)
	-
	\sum_{i<j}\sum_{q,\ell}
	\tau_{iq}\tau_{j\ell}
	\Bigg[
	\rho_{q\ell}^{\,i,j}\frac{K_{ij}^2}{2\sigma_1^2}
	+
	(1-\rho_{q\ell}^{\,i,j})\,\xi_0 |K_{ij}|
	\Bigg]
	-
	\sigma_1 \sum_{i=1}^p K_{ii},
\end{equation}
where $\ell(K;S) = \frac{n}{2}\bigl(\log\det K-\tr(SK)\bigr)$ is, up to an additive constant, the Gaussian log-likelihood based on the sample covariance matrix $S$.
As direct maximization is involved, we follow  \citet{Meinshausen2006} and \citet{Chiquet2011} and adopt a   pseudo-likelihood approach that splits the optimization problem into $p$ independent regression problems. To state the pseudo likelihood, for each \(i\in\{1,\ldots,p\}\), we denote  \(\mathbf X_i\in\mathbb R^n\) the \(i\)th column of the data matrix \(\mathbf X\) and  \(\mathbf X_{-i}\in\mathbb R^{n\times(p-1)}\)  the matrix obtained by removing that column.    Then the conditional distribution of  $\mathbf X_i$  given the other variables is Gaussian with
\[
\mathbf X_i \mid \mathbf X_{-i}
\sim
\mathcal N\!\left(\mathbf X_{-i}\beta_i,\ \sigma_i^2 I_n\right),
\qquad
\sigma_i^2=\frac{1}{K_{ii}},
\qquad
\beta_i=-\frac{K_{-i,i}}{K_{ii}}\in\mathbb R^{p-1},
\]
where \(K_{-i,i}\in\mathbb R^{p-1}\) denotes the \(i\)th column of \(K\) with its diagonal entry removed. We introduce the  pseudo log-likelihood 
\begin{equation}
	\check\ell(K;S)=
	\sum_{i=1}^p \log p(\mathbf X_i \mid \mathbf X_{-i}, K)
	=
	\sum_{i=1}^p
	\left[
	-\frac{n}{2}\log(2\pi)
	-\frac{n}{2}\log\sigma_i^2
	-\frac{1}{2\sigma_i^2}\|\mathbf X_i-\mathbf X_{-i}\beta_i\|_2^2
	\right].
	\label{eq:pseudologlik_revised}
\end{equation}
Finally, in \eqref{eq:evidence}, we replace $\ell(K;S)$ with $\check\ell(K;S)$ leading to an optimization problem that amounts to solving \(p\) separate regression problems of dimension \(p-1\). 
More precisely,
to update  the \(i\)th column of \(K\)  we solve
\begin{equation}
	\widehat\beta_i
	=
	\arg\min_{\beta\in\mathbb R^{p-1}}
	\left\{
	\frac{1}{2\sigma_i^2}\|\mathbf X_i-\mathbf X_{-i}\beta\|_2^2
	+
	\sum_{j\neq i}P_{1,ij}|\beta_{ij}|
	+
	\sum_{j\neq i}P_{2,ij}\beta_{ij}^2
	\right\},
	\label{eq:opt_column}
\end{equation}
with weights
\begin{equation}
	P_{1,ij}
	=
	\xi_0(1-p_{ij})|K_{ii}|,
	\qquad
	P_{2,ij}
	=
	\frac{1}{2\sigma_1^2}p_{ij}K_{ii}^2,
	\qquad
	p_{ij}
	=
	\sum_{q,\ell}\tau_{iq}\tau_{j\ell}\rho_{q\ell}^{\,i,j}.
	\label{eq:weights_beta}
\end{equation}
The problem in \eqref{eq:opt_column}  corresponds to a weighted elastic-net regression, where the variational posterior edge probabilities $p_{ij}$ modulate the weighting of the Lasso and Ridge penalties. This induces an adaptive trade-off between sparsity and smooth shrinkage. In particular, edges with high posterior probability are mainly governed by ridge-type penalization, whereas edges with low posterior probability are mainly driven by Lasso-type shrinkage.
The criterion in \eqref{eq:opt_column} is convex and becomes strictly convex as soon as $P_{2,ij} > 0$ for all $j \neq i$. Therefore, each nodewise subproblem is well-posed and admits a unique minimizer whenever $p_{ij} > 0$ for all $j \neq i$.

Once \(\widehat\beta_i\) has been obtained, the corresponding entries of the precision matrix are updated as
\begin{equation}
	\widehat K_{ii}
	=
	\frac{1}{\widehat\sigma_i^2}
	=
	\frac{n}{\|\mathbf X_i-\mathbf X_{-i}\widehat\beta_i\|_2^2},
	\qquad
	\widehat K_{ji}
	=
	-\,\widehat\beta_{ij}\,\widehat K_{ii},
	\qquad j\neq i.
	\label{eq:K_update_column}
\end{equation}
Since the nodewise procedure generally yields a non-symmetric matrix, we symmetrize \(\widehat K\) using the OR rule: for each pair \((i,j)\), we retain the coefficient with the largest absolute value, together with its sign, namely
\begin{equation}
	\widehat K_{ij}^{\,\mathrm{sym}}
	=
	\widehat K_{ji}^{\,\mathrm{sym}}
	=
	\begin{cases}
		\widehat K_{ij}, & \text{if } |\widehat K_{ij}| \ge |\widehat K_{ji}|,\\
		\widehat K_{ji}, & \text{otherwise.}
	\end{cases}
	\label{eq:OR_rule}
\end{equation}
As our primary objective is graph recovery rather than precision matrix estimation, global positive-definiteness of \(\widehat K\) is not enforced at this stage.

Algorithm~\ref{algo:K_update} summarizes the precision matrix update, while Algorithm~\ref{algo:VEM} describes the full inference algorithm alternating between   VEM updates  and   precision matrix updates.

\begin{algorithm}[t]
	\caption{Adaptive elastic-net estimation of the precision matrix}
	\label{algo:K_update}
	\KwIn{Data matrix \(\mathbf X\),  variational parameters \((\tau,\rho)\), hyperparameters \((\sigma_1,\xi_0)\), current precision matrix  $K$}
	\KwOut{Updated precision matrix \(K\)}
	
	\For{ each node \(i=1,\dots,p\)}{
		Compute the variational posterior probability of edge presence $p_{ij}$ for $j\neq i$ with \eqref{eq:weights_edge};\\
		Compute the edge-penalty weights $P_{1,ij}$ and $P_{2,ij}$ for $j\neq i$ with \eqref{eq:weights_beta};\\
		Solve the weighted elastic-net problem \eqref{eq:opt_column} to obtain $\widehat\beta_i\ \in \mathbb{R}^{p-1}$;\\
		Update the $i$-th column of $K$ with \eqref{eq:K_update_column}; \\
	}
	Symmetrise the matrix $K$ using the OR rule \eqref{eq:OR_rule}
\end{algorithm}

\begin{algorithm}[t]
	\caption{Alternating inference algorithm}
	\label{algo:VEM}
	\KwIn{Data matrix \(\mathbf X\), number of blocks \(Q\), hyperparameters \((\xi_0,\sigma_1)\)}
	\KwOut{Estimates of $K, \tau, \rho, \theta$} 
\textbf{Initialization}\;
Initialize  $K^{(0)}, \tau^{(0)}, \theta^{(0)}$ and $t=0$\\
Compute $\rho^{(0)}$ from $K^{(0)}$ and $\theta^{(0)}$ with equation \eqref{eq:rho_update}\\
\While{not converged}{
	Set $t=t+1$;\\
	Update the precision matrix $K^{(t)}$ using Algorithm~	\ref{algo:K_update} from variational parameter $(\tau^{(t-1)}, \rho^{(t-1)})$ and current  $K^{(t-1)}$;\\
	Update the varitional parameters $\tau^{(t)}, \rho^{(t)}$ and model parameter $\theta^{(t)}$ using the VEM algorithm detailled in Appendix~\ref{app:vem-derivations} from $K^{(t)}$ and current $(\tau^{(t-1)}, \rho^{(t-1)}, \theta^{(t-1)})$ \\
}
\end{algorithm}

\subsection{Graph selection by multiple testing}
\label{subsec:graph_inference}

We now address graph inference through the multiple testing problem given in \eqref{eq:hypothesis}, where our goal is to decide whether an edge is present.
We propose a multiple testing procedure based on the  $\ell$-values introduced by \citet{Efron2001,Efron2004} and defined as the  conditional posterior probabilities
\begin{equation}
	\ell_{ij}(K,Z,\theta)
	= 
	\pr(A_{ij}=0\mid K_{ij},Z_i,Z_j; \theta),\quad (i,j) \in \mathcal{A}.
	\label{eq:lvalue_def}
\end{equation}
In the spike-and-slab model, this quantity has the explicit form
\begin{equation}
	\ell_{ij}(K,Z,\theta)
	=1- \rho_{Z_i Z_j}^{ij}=
	\frac{(1-\omega_{Z_iZ_j})\,g_{\xi_0}(K_{ij})}
	{(1-\omega_{Z_iZ_j})\,g_{\xi_0}(K_{ij})
		+\omega_{Z_iZ_j}\,\phi(K_{ij};0,\sigma_1^2)}.
	\label{eq:lvalue_oracle}
\end{equation}
Intuitively, small values of \(\ell_{ij}\) provide evidence against \(H_{0,ij}\), and therefore in favour of the presence of an edge.  A multiple testing procedure consists in thresholding the \(\ell\)-values at an appropriate level \(t\) chosen to control the FDR defined as
\begin{equation}
	\mathrm{FDR}_{\theta}(t)
	=
	\mathbb E_{\theta}\!\left[
	\frac{
		\sum_{(i,j)\in\mathcal A}(1-A_{ij})\,\mathds{1}_{\{\ell_{ij}(K,Z,\theta)\le t\}}
	}{
		\sum_{(i,j)\in\mathcal A}\mathds{1}_{\{\ell_{ij}(K,Z,\theta)\le t\}}\vee 1
	}
	\right].
	\label{eq:fdr_definition}
\end{equation}. 
Here \(\mathbb E_{\theta}\) denotes expectation under the model indexed by \(\theta\).

In multiple testing literature, the marginal false discovery rate (MFDR) is often used as a convenient substitute for the FDR. The MFDR is defined as
\begin{equation}
	\mathrm{MFDR}_{\theta}(t)
	=
	\frac{
		\mathbb E_{\theta}\!\left[
		\sum_{(i,j)\in\mathcal A}(1-A_{ij})\,\mathds{1}_{\{\ell_{ij}(K,Z,\theta)\le t\}}
		\right]
	}{
		\mathbb E_{\theta}\!\left[
		\sum_{(i,j)\in\mathcal A}\mathds{1}_{\{\ell_{ij}(K,Z,\theta)\le t\}}
		\right]
	},
	\label{eq:mfdr_definition}
\end{equation}
with the convention \(0/0=0\). 
We reject the \(H_{0,ij}\) whenever $ \ell_{ij} \le t_{\alpha}$ with $t_\alpha$ such that $\mathrm{MFDR}_{\theta}(t_{\alpha})=\alpha$. To circumvent the explicit calculation of this threshold, we use the following quantities 
\begin{equation}
	q_{ij}(K,Z,\theta)
	=
	\mathrm{MFDR}_{\theta}\big(\ell_{ij}(K,Z,\theta)\big),
	\qquad
	(i,j)\in\mathcal A,
	\label{eq:qvalues_oracle}
\end{equation}
These quantitites, called $q$-values, were used in \citet{Rebafka2022} and can be interpreted as adjusted \(\ell\)-values in the sense that they can be compared directly with the target FDR level \(\alpha\).

The $q$-values admit an explicit expression in our model, as detailed in Appendix~\ref{app:proof_qvalues}. 

Using the estimates of \(K\), \(Z\), and \(\theta\) obtained from Algorithm~\ref{algo:VEM}, we define the empirical \(q\)-values as
\begin{equation}
	\widehat q_{ij}
	=
	q_{ij}(\widehat K,\widehat Z;\widehat\theta),
	\qquad
	(i,j)\in\mathcal A,
	\label{eq:qvalues_plugin}
\end{equation}
where the node clustering \(\widehat Z\) is obtained from the variational parameters through
\begin{equation}
	\widehat Z_i=\arg\max_{1\le q\le Q}\widehat\tau_{iq},
	\qquad i=1,\dots,p.
	\label{eq:Zhat_from_tau}
\end{equation}

Then, the final estimated graph at nominal level \(\alpha\) is
\begin{equation}
	\widehat A^\alpha_{ij}
	=
	\mathds{1}_{\{\widehat q_{ij}\le \alpha\}},
	\qquad
	(i,j)\in\mathcal A.
	\label{eq:graph_decision_rule}
\end{equation}

\subsection{Calibration of hyperparameters and selection of the number of blocks}
\label{subsec:hyperparam_calibration}

The spike-and-slab hyperparameters \((\xi_0,\sigma_1)\) and the number of blocks \(Q\) are calibrated jointly using an appropriate Bayesian information criterion. The choice of \((\xi_0,\sigma_1)\) is guided by the theoretical bounds established in Section~\ref{subsec:error_bounds}. In particular, the theory requires the spike level \(\xi_0\) to be of order \(\sqrt{n\log p}\), while the slab variance \(\sigma_1^2\) remains in a bounded range. This motivates the calibration scheme
\[
\xi_0=c\,\sqrt{n\log p},
\qquad c\in\mathcal C,
\]
for a finite grid \(\mathcal C\) of positive constants, together with a finite grid
\[
\sigma_1\in\mathcal S\subset[\underline{\sigma},\overline{\sigma}],
\]
where \(0<\underline{\sigma}<\overline{\sigma}<\infty\) are fixed. Thus, the theoretical analysis does not determine a unique pair \((\xi_0,\sigma_1)\), but restricts calibration to a theoretically admissible low-dimensional family. For each \(Q\in\mathcal Q=\{Q_{\min},\dots,Q_{\max}\}\), for a fixed nominal level \(\alpha\), and for each pair \((\xi_0,\sigma_1)\in\mathcal C\times\mathcal S\), we run the full graph inference procedure and obtain a graph estimate \(\widehat A_{\alpha}^{(Q,\xi_0,\sigma_1)}\). These candidate graphs are then compared through a Bayesian information criterion,
\begin{equation}
	\mathrm{BIC}(\xi_0,\sigma_1;Q,\alpha)
	=
	-2\,\widetilde\ell\!\left(X;\widehat A_{\alpha}^{(Q,\xi_0,\sigma_1)}\right)
	+
	\log n\,\sum_{i<j}\widehat A_{\alpha,ij}^{(Q,\xi_0,\sigma_1)},
	\label{eq:bic_pseudolik}
\end{equation}
where \(\widetilde\ell(X;A)\) denotes the Gaussian pseudo-loglikelihood associated with the candidate adjacency matrix \(A\). The second term penalizes model complexity through the number of selected edges. For fixed \(Q\), the selected hyperparameters are
\[
(\widehat\xi_0^{(Q)},\widehat\sigma_1^{(Q)})
\in
\arg\min_{(\xi_0,\sigma_1)\in\mathcal C\times\mathcal S}
\mathrm{BIC}(\xi_0,\sigma_1;Q,\alpha).
\]
Then, for the selection of the number of \(Q\), we define the criterion
\begin{equation}
	\mathrm{BIC}_Q
	=
	-2\,\widetilde\ell\!\left(
	X;\widehat A_{\alpha}^{(Q,\widehat\xi_0^{(Q)},\widehat\sigma_1^{(Q)})}
	\right)
	+
	(Q-1)\log p
	+
	\frac{Q(Q+1)}{2}\log\!\left(\frac{p(p-1)}{2}\right),
	\label{eq:BIC_Q}
\end{equation}
where the penalty accounts for the complexity of the block model: the block proportions \(\pi=(\pi_1,\dots,\pi_Q)\) contribute \(Q-1\) free parameters, and the symmetric connectivity matrix \(\omega\in(0,1)^{Q\times Q}\) contributes \(Q(Q+1)/2\) distinct entries. Finally, the selected number of blocks is
\[
\widehat Q
\in
\arg\min_{Q\in\mathcal Q}\mathrm{BIC}_Q.
\]

The complete calibration and model selection procedure is summarized in Algorithm~\ref{algo:HSSGGM_complete} in the Appendix~\ref{app:calibration}.

\section{Theoretical results}

\subsection{Error bounds for the adaptive elastic-net estimator}
\label{subsec:error_bounds}

We study the statistical properties of the nodewise adaptive elastic-net
estimator used in the M-step of the VEM algorithm. At each iteration, the
variational parameters \(\tau\) and \(\rho\) are fixed from the preceding E-step,
so that the quantities
\[
p_{ij} = \sum_{q,\ell} \tau_{iq}\tau_{j\ell}\rho_{q\ell}^{\,i,j}
\]
are treated as deterministic. The analysis in this section is therefore
conditional on these weights.

Fix a node \(i \in \{1,\ldots,p\}\) and consider the Gaussian nodewise regression
model
\begin{equation}
	\mathbf X_i = \mathbf X_{-i}\beta_i^\star + \varepsilon_i,
	\qquad
	\varepsilon_i \sim \mathcal{N}(0,\sigma_i^2 I_n),
	\label{eq:nodewise_model}
\end{equation}
where \(\beta_i^\star \in \mathbb{R}^{p-1}\) denotes the true regression vector,
and \(K_{ii}\) is treated as known and fixed. Let
\[
S_i = \{j \neq i : \beta_{ij}^\star \neq 0\},
\qquad
s_i = |S_i|,
\]
denote the support of \(\beta_i^\star\) and its cardinality. The estimator
\(\widehat\beta_i\) is defined by the nodewise adaptive elastic-net problem
\eqref{eq:opt_column}, with weights given in \eqref{eq:weights_beta}. We impose
the following assumptions.

\begin{assumption}[Sub-Gaussian design]
	\label{ass:error_bound_A1}
	For each node \(i\), the rows of \(\mathbf X_{-i}\in\mathbb R^{n\times(p-1)}\) are
	independent copies of a centered random vector in \(\mathbb R^{p-1}\) with
	covariance matrix \(\Sigma_{-i,-i}\). Moreover, there exists a constant
	\(\sigma_X>0\), uniform in \(n\) and \(p\), such that for every unit vector
	\(v\in\mathbb R^{p-1}\),
	\[
	\mathbb E\!\left[\exp\!\bigl(t\langle \mathbf X_{-i}^{(1)},v\rangle\bigr)\right]
	\le
	\exp\!\left(\frac{t^2\sigma_X^2}{2}\right),
	\qquad \forall\, t\in\mathbb R,
	\]
	where \(\mathbf X_{-i}^{(1)}\) denotes a generic row of \(\mathbf X_{-i}\). In addition,
	the eigenvalues of \(\Sigma_{-i,-i}\) are uniformly bounded:
	\[
	0<\underline\kappa \le \lambda_{\min}(\Sigma_{-i,-i})
	\le \lambda_{\max}(\Sigma_{-i,-i}) \le \overline\kappa < \infty.
	\]
\end{assumption}

\begin{assumption}[Sparsity and bounded signal]
	\label{ass:error_bound_A2}
	For each node \(i\), the true coefficient vector \(\beta_i^\star\) is sparse
	with support \(S_i\), and satisfies
	\[
	\max_{j \in S_i} |\beta_{ij}^\star| \le B
	\]
	for some constant \(B > 0\), uniform in \(n\) and \(p\).
\end{assumption}

\begin{assumption}[Separation of adaptive weights]
	\label{ass:error_bound_A3}
	Define
	\[
	\overline{p}_i = \max_{j \in S_i^c} p_{ij},
	\qquad
	\underline{p}_i = \min_{j \in S_i} p_{ij}.
	\]
	Assume that \(\overline{p}_i < 1/2\).
\end{assumption}

\begin{assumption}[Compatibility condition]
	\label{ass:error_bound_A4}
	Let
	\[
	L_i = \frac{5-2\underline p_i}{1-2\overline p_i}.
	\]
	There exists a constant \(\phi_i > 0\), uniform in \(n\) and \(p\), such that
	for every \(\delta \in \mathbb{R}^{p-1}\) satisfying
	\(\|\delta_{S_i^c}\|_1 \le L_i \|\delta_{S_i}\|_1\),
	\begin{equation}
		\|\mathbf X_{-i}\delta\|_2^2
		\ge
		\frac{n\phi_i^2}{s_i}\|\delta_{S_i}\|_1^2.
		\label{eq:compatibility_main}
	\end{equation}
\end{assumption}

Assumption~\ref{ass:error_bound_A1} is a standard regularity condition on the
design matrix and is automatically satisfied in the Gaussian graphical model
considered here. Assumption~\ref{ass:error_bound_A2} is the usual sparsity and
bounded-signal condition. Assumption~\ref{ass:error_bound_A3} ensures that inactive
coordinates receive stronger \(\ell_1\)-penalization than active ones.
Assumption~\ref{ass:error_bound_A4} is a compatibility condition adapted to the cone
constraint induced by the adaptive elastic-net; see, for instance,
\citet{Meinshausen2006,Rothman2008,Ravikumar2011,Cai2011,Loh2011} for related
assumptions in the analysis of \(\ell_1\)-regularized estimators.

Recall that \(\xi_0\) controls the concentration of the Laplace spike around
zero in the prior. The next result shows that taking \(\xi_0\) of order
\(\sqrt{n\log p}\) yields the usual high-dimensional \(\ell_1\)-error rate.

\begin{theorem}[Non-asymptotic \(\ell_1\)-error bound]
	\label{thm:error_bound_weighted_enet}
	Under Assumptions~\ref{ass:error_bound_A1}--\ref{ass:error_bound_A4}, conditionally
	on the weights \((p_{ij})\) fixed from the preceding E-step, and provided that
	\[
	\xi_0 \ge C\sqrt{n\log p}
	\qquad\text{and}\qquad
	\xi_0 \ge \frac{B K_{ii}}{\sigma_1^2},
	\]
	for a sufficiently large constant \(C>0\), the nodewise adaptive elastic-net
	estimator satisfies, with probability at least \(1-2p^{-c}\),
	\[
	\|\widehat\beta_i-\beta_i^\star\|_1
	\le
	C_i\,\frac{s_i\xi_0}{n},
	\]
	where \(C_i>0\) depends only on \(\phi_i\), \(\underline p_i\), and
	\(\overline p_i\).
\end{theorem}

Under the calibration \(\xi_0 \asymp \sqrt{n\log p}\),
Theorem~\ref{thm:error_bound_weighted_enet} yields
\[
\|\widehat\beta_i - \beta_i^\star\|_1
=
O_{\mathbb P}\!\left(s_i\sqrt{\frac{\log p}{n}}\right).
\]
The proof is given in Appendix~\ref{sec:proof_error_bound}. The lower bound
\(\xi_0 \ge C\sqrt{n\log p}\) comes from the concentration of the stochastic
term \(n^{-1}\mathbf X_{-i}^\top \varepsilon_i\). This is the standard minimax-optimal \(\ell_1\)-rate for sparse linear regression
in high dimension \citep{Raskutti2009}. The adaptive weights do not alter the
rate itself, but affect the constant through \(\underline p_i\) and
\(\overline p_i\): better separation between active and inactive weights leads to
a tighter bound. Thus, the role of the adaptive elastic-net is not to improve
the estimation rate, but to concentrate regularization on the most relevant
coordinates.

\section{Numerical experiments}

We assess the empirical performance of the proposed procedure on synthetic data and on a breast cancer gene expression dataset.

\subsection{Synthetic data}
\label{subsec:synthetic_data}

\begin{figure}[!htbp]
\centering
\begin{subfigure}{0.22\textwidth}
\centering
\includegraphics[width=\textwidth]{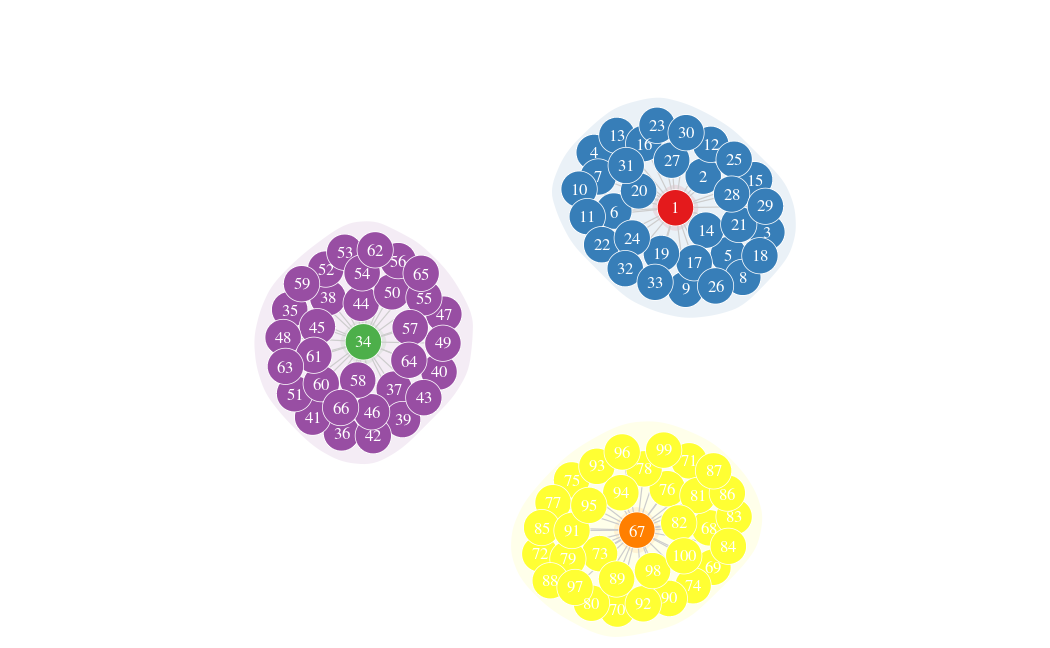}
\caption{Hub}
\end{subfigure}
\hfill
\begin{subfigure}{0.22\textwidth}
\centering
\includegraphics[width=\textwidth]{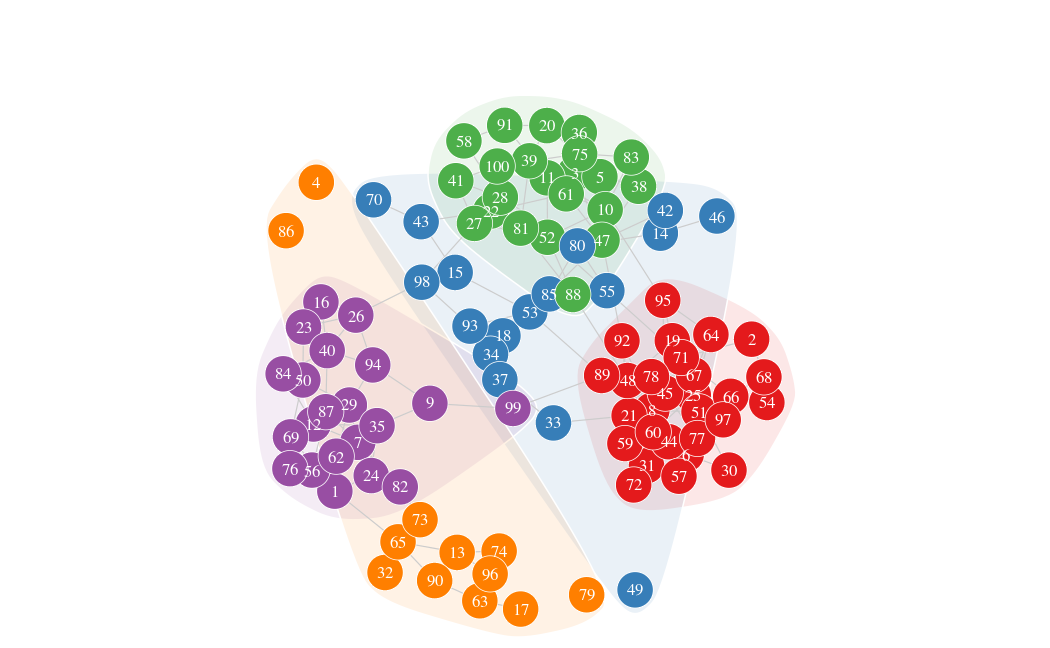}
\caption{SBM}
\end{subfigure}
\hfill
\begin{subfigure}{0.22\textwidth}
\centering
\includegraphics[width=\textwidth]{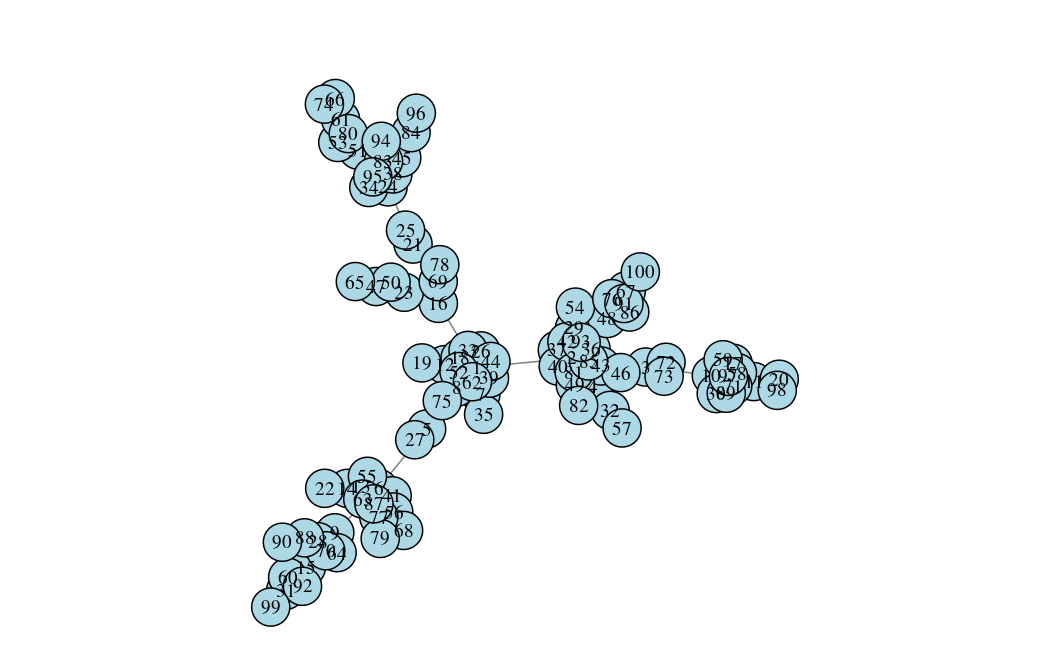}
\caption{Scale-free}
\end{subfigure}
\hfill
\begin{subfigure}{0.22\textwidth}
\centering
\includegraphics[width=\textwidth]{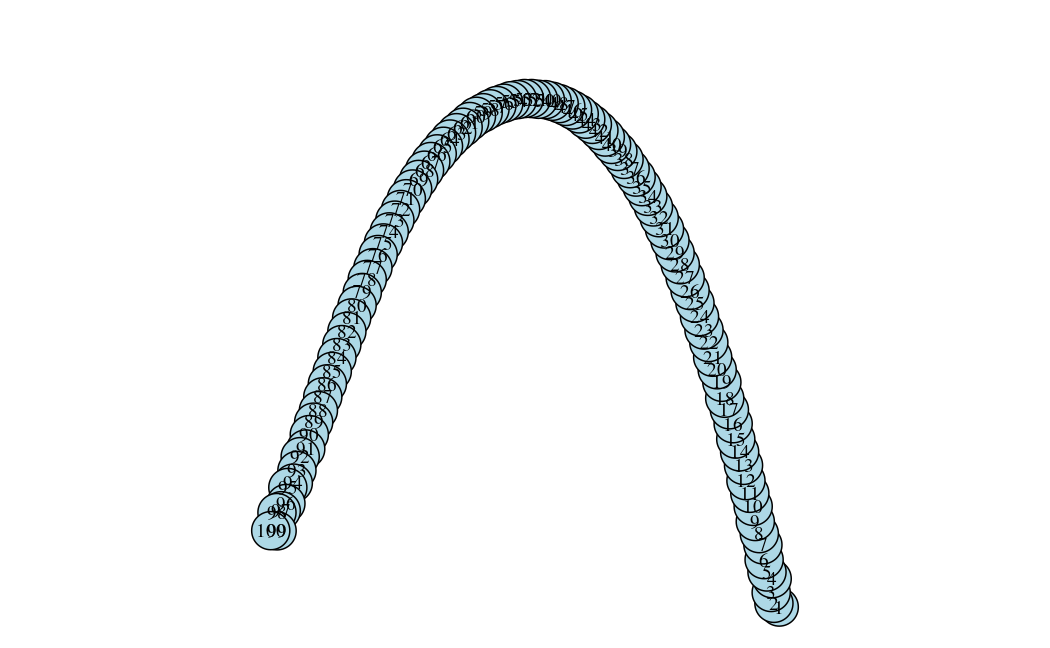}
\caption{Band}
\end{subfigure}
\caption{Illustration of the four graph structures considered in the simulation study.}
\label{fig:graph_topology}
\end{figure}

We assess the performance of the proposed HSS-GGM procedure on synthetic datasets for which the underlying graph structure is known. Each dataset consists of \(n\) independent observations drawn from a multivariate Gaussian distribution with mean zero and precision matrix \(K_0\), whose sparsity pattern is encoded by a binary adjacency matrix \(A_0\). To cover a range of realistic dependence structures, we consider four graph topologies, illustrated in Figure~\ref{fig:graph_topology}: a three-hub graph, a stochastic block model graph, a scale-free graph and a band graph.

The three-hub graph contains three highly connected hub nodes linked to many peripheral nodes. The stochastic block model graph has higher connection probabilities within groups than between groups. The scale-free graph has a heavy-tailed degree distribution, with a few highly connected nodes and many nodes of low degree. The band graph has a banded adjacency matrix and is representative of ordered dependence structures such as autoregressive models.

For each topology, we follow \citet{Kilian2024} and the \texttt{huge} package to generate a positive-definite precision matrix \(K_0\) from the binary adjacency matrix \(A_0\). Specifically, we define
\[
K_0
=
\gamma A_0
+
\bigl(\bigl|\lambda_{\min}(\gamma A_0)\bigr|+\beta\bigr) I_p,
\]
with \(\gamma=0.3\) and \(\beta=0.2\). The corresponding covariance matrix is \(\Sigma_0=K_0^{-1}\), from which we draw \(n\) independent observations from \(\mathcal{N}(0,\Sigma_0)\).

In the main text we consider the high-dimensional setting \(n=p=100\). For each topology, we generate 50 independent datasets, and all results are aggregated across replications.

We compare HSS-GGM with six competing methods: Meinshausen--Bühlmann neighbourhood selection \citep{Meinshausen2006} and graphical lasso \citep{Friedman2008}, both implemented through the \texttt{huge} package \citep{Zhao2012}; SIMoNe \citep{Ambroise2008}; GFC-L \citep{Liu2013} through the \texttt{SILGGM} package \citep{Zhang2018}; BAGUS \citep{Gan2019}; and NSBM-GGM \citep{Kilian2024}, implemented in \texttt{noisysbmGGM}.

Graph recovery is evaluated by comparing the estimated adjacency matrix \(\widehat A\) with the true adjacency matrix \(A_0\). We report the false discovery proportion,
\[
\FDP
=
\frac{|\{(i,j): \widehat A_{ij}=1,\ A_{0,ij}=0\}|}{|\{(i,j): \widehat A_{ij}=1\}|\vee 1},
\]
and the true discovery proportion,
\[
\TDP
=
\frac{|\{(i,j): \widehat A_{ij}=1,\ A_{0,ij}=1\}|}{|\{(i,j): A_{0,ij}=1\}|}.
\]

For procedures explicitly designed to control the false discovery rate, namely HSS-GGM, GFC-L and NSBM-GGM, we set the nominal level to \(\alpha=0.1\). For BAGUS, we follow \citet{Gan2019} and recover the graph by thresholding posterior edge probabilities at \(1/2\). For graphical lasso and Meinshausen--Bühlmann neighbourhood selection, we use the graph selected by the \texttt{huge} implementation. For SIMoNe, the built-in AIC or BIC often selects empty or nearly empty graphs in our experiments. We therefore retain along the regularization path a sparse graph containing about \(3\%\) of the possible edges. This level was chosen because it is of the same order as that returned by the other procedures in the settings considered.

\begin{figure}[!htbp]
\centering
\includegraphics[width=\textwidth]{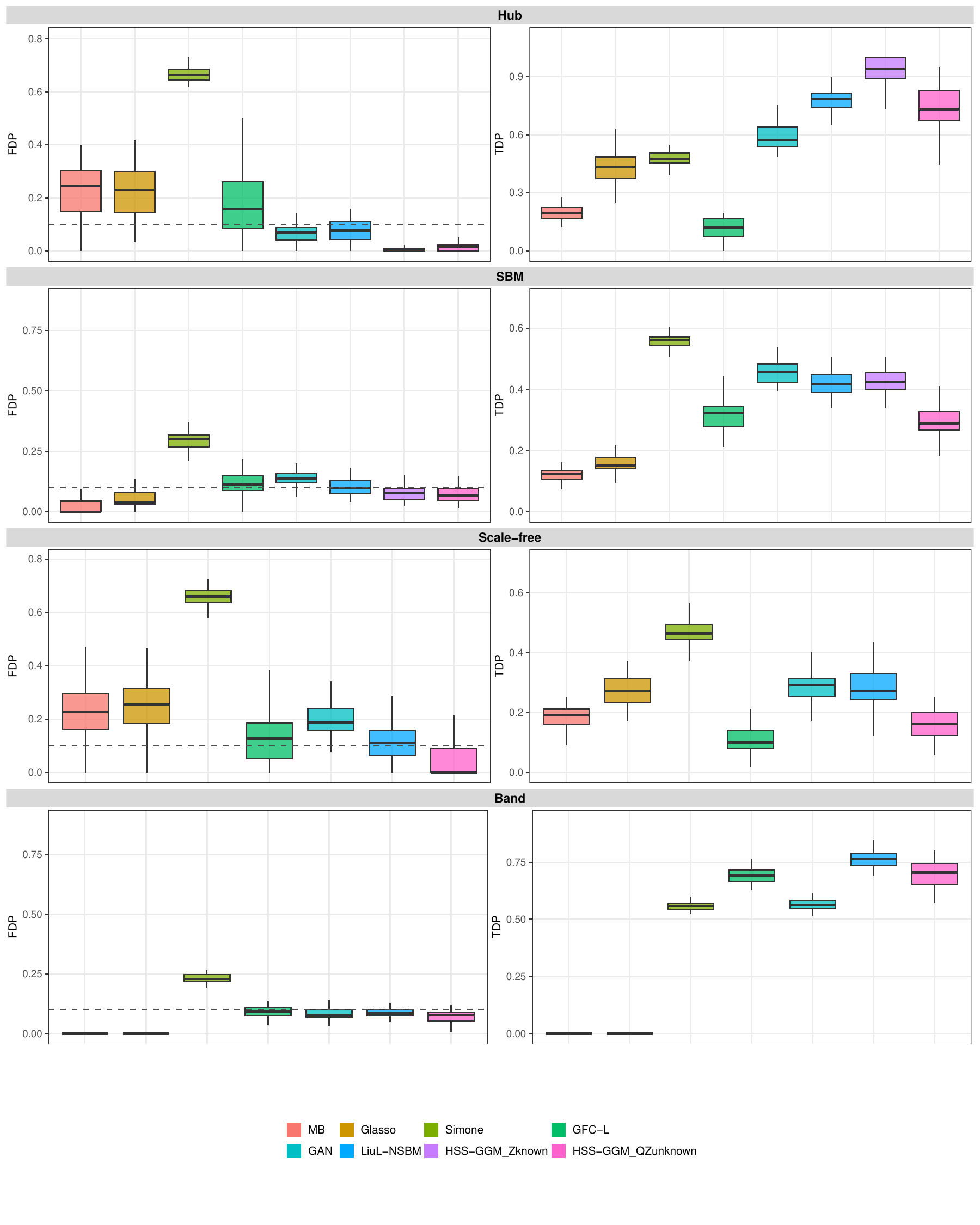}
\caption{Empirical \FDP\ and \TDP\ distributions for the four graph topologies when \(n=p=100\). For each method, boxplots summarize performance over 50 replications. The horizontal dashed line indicates the nominal false discovery rate level \(\alpha=0.1\).}
\label{fig:main_simu_results_p100}
\end{figure}

Figure~\ref{fig:main_simu_results_p100} summarizes the empirical \FDP\ and \TDP. We first consider false discovery control, which is the primary objective of the proposed procedure. In all settings, the empirical \FDP\ of HSS-GGM remains below the nominal level \(\alpha=0.1\). By contrast, several competing procedures exceed the target level in topologies with marked degree heterogeneity, especially hub and scale-free graphs. Although most of these methods are not designed to control the false discovery rate, the observed inflation illustrates the risk of false positives when they are used for graph inference. Even among procedures explicitly targeting false discovery rate control, namely GFC-L and NSBM-GGM, the empirical \FDP\ occasionally exceeds the nominal level in these more difficult settings.

We next consider statistical power. In structured settings such as stochastic block model and hub graphs, and especially when group memberships are known, HSS-GGM achieves the largest \TDP, including relative to methods that do not aim to control the false discovery rate and tend to overselect edges. This illustrates the benefit of incorporating group structure into the shrinkage mechanism. Across all graph topologies, HSS-GGM achieves competitive or superior power while maintaining reliable false discovery control. In hub graphs this gain is particularly marked. Under structural misspecification, such as in band graphs, HSS-GGM remains competitive and does not show substantial performance loss, indicating robustness beyond the assumed block structure. Additional simulation results are reported in Appendix~\ref{app:additional_simulations}.

\subsection{Breast cancer gene expression network}

We apply HSS-GGM to the breast cancer gene expression dataset of \citet{Hess2006}, which was also used in the development of SIMoNe \citep{Ambroise2008}. The dataset contains expression profiles from 133 patients with stage I--III breast cancer treated with neoadjuvant chemotherapy. Clinical response is classified as either pathologic complete response (pCR; 34 patients) or residual disease (not-pCR; 99 patients). Following \citet{Hess2006} and \citet{Natowicz2008}, who identified a 26-gene signature with strong predictive value for treatment outcome, we restrict attention to this curated gene set, so that \(p=26\).

Because gene expression profiles may differ substantially between responders and non-responders, the assumption of identically distributed observations is unlikely to hold across the full cohort. We therefore focus on the pCR subgroup and estimate the corresponding conditional independence network using HSS-GGM at nominal level \(\alpha=0.05\).

To compare the inferred graph with those obtained by competing procedures, we consider the same methods as in the simulation study. For procedures explicitly designed to control the false discovery rate, namely HSS-GGM, GFC-L and NSBM-GGM, we use the nominal level \(\alpha=0.05\). For BAGUS, we recover the graph by thresholding posterior edge probabilities at \(1/2\), following \citet{Gan2019}. For graphical lasso and Meinshausen--Bühlmann neighbourhood selection, we use the graph selected by the \texttt{huge} implementation. For SIMoNe, the built-in AIC or BIC may lead to empty or nearly empty graphs; we therefore retain along the regularization path a sparse graph with edge density comparable to that returned by the other procedures.

\begin{figure}[!htbp]
\centering
\begin{subfigure}[t]{0.48\textwidth}
\centering
\includegraphics[width=\textwidth]{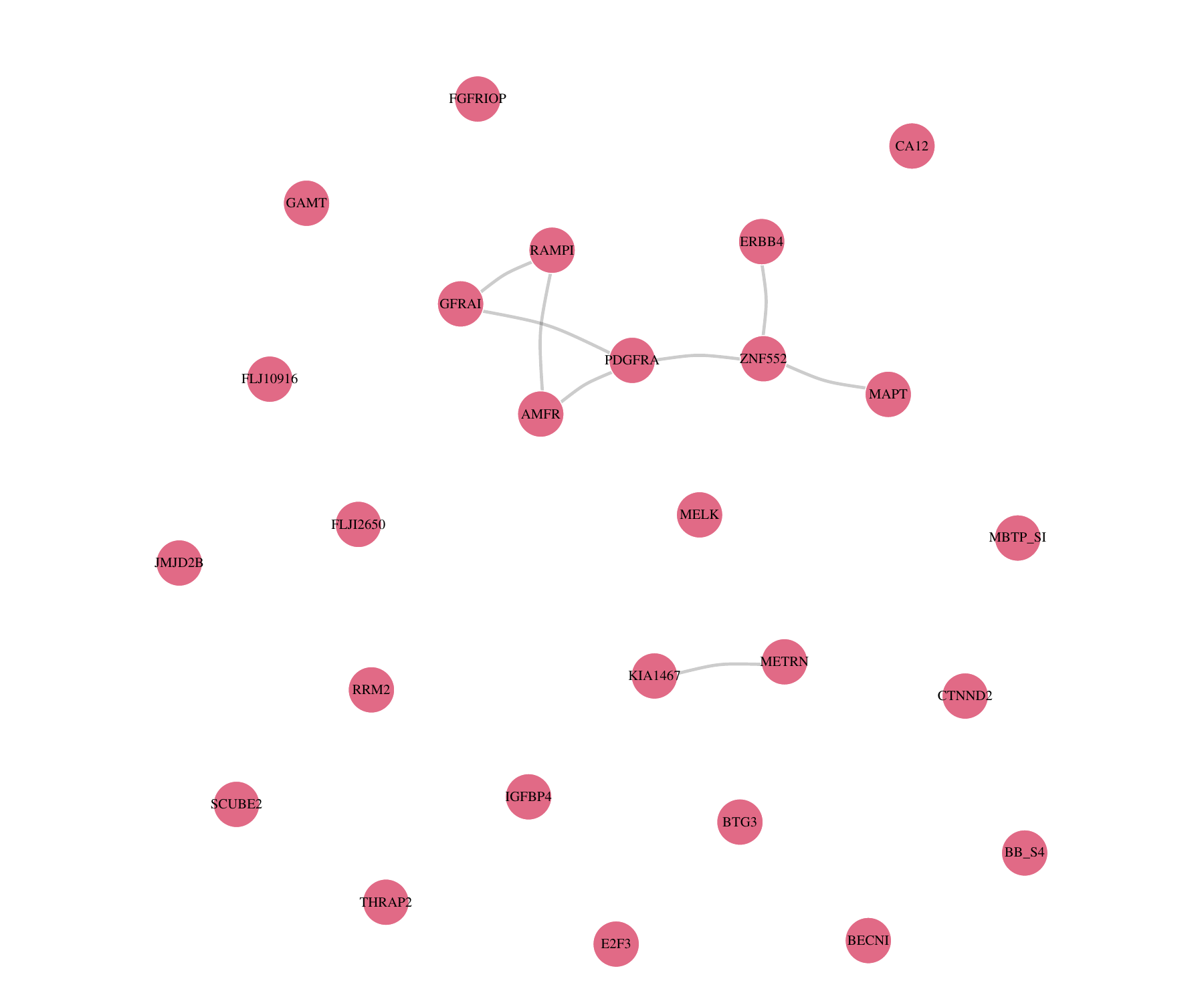}
\caption{Network inferred by HSS-GGM for the pCR subgroup}
\end{subfigure}
\hfill
\begin{subfigure}[t]{0.48\textwidth}
\centering
\includegraphics[width=\textwidth]{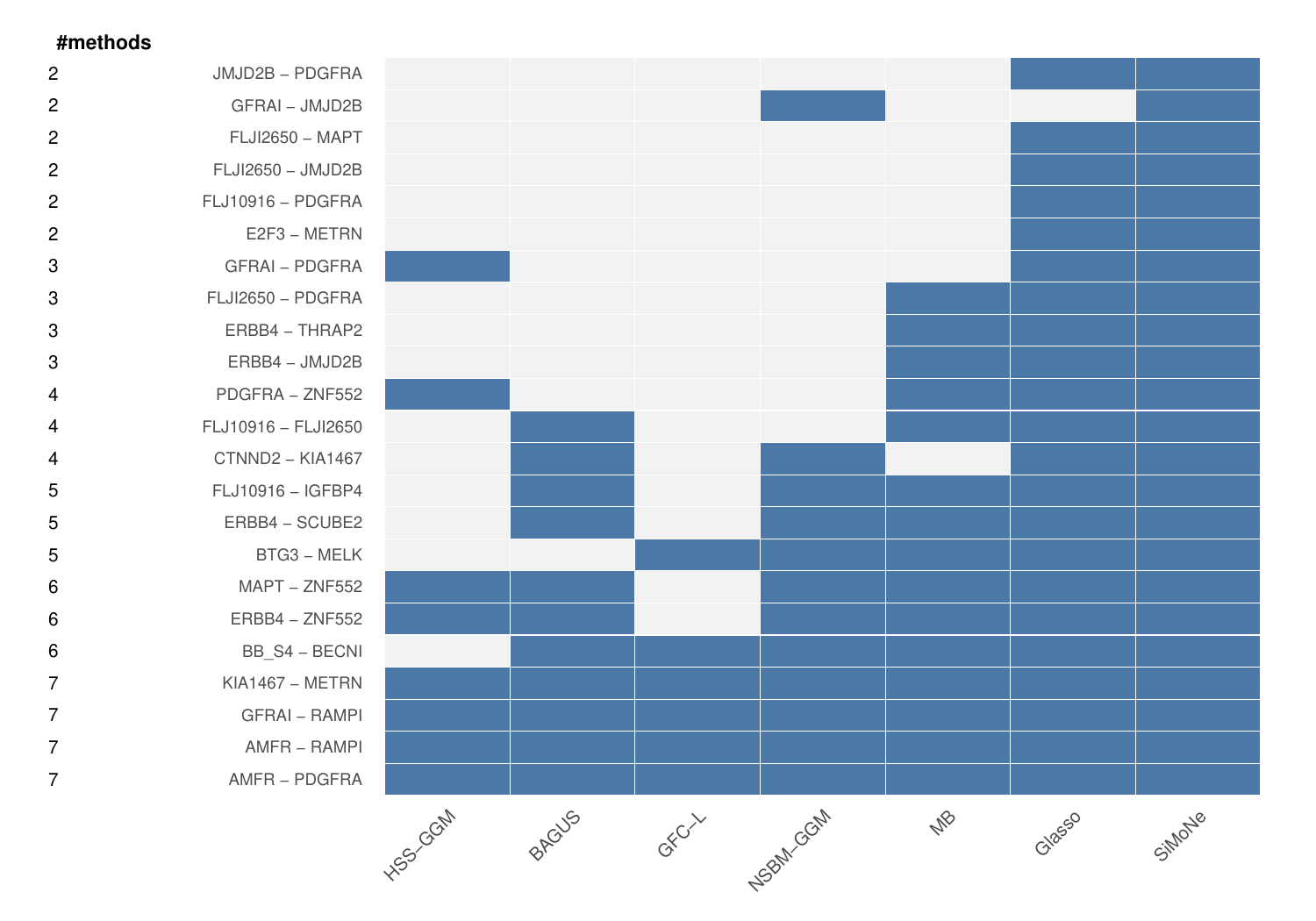}
\caption{Edge detection across methods}
\end{subfigure}
\caption{(a) Conditional independence network inferred by HSS-GGM at nominal level \(\alpha=0.05\) for the pCR subgroup. (b) Comparison of edges detected by different network inference methods. The left column reports the number of methods detecting each interaction.}
\label{fig:cancer_graph}
\end{figure}

Figure~\ref{fig:cancer_graph} displays the conditional independence network inferred by HSS-GGM for the pCR subgroup, together with a summary of edge detection across methods. HSS-GGM identifies 8 conditional dependencies among the 26 genes. Each of these edges is also recovered by at least one competing method, indicating substantial agreement across procedures on the most stable interactions.

The inferred network is relatively sparse and overlaps substantially with the graphs obtained by the alternative methods. This is consistent with the simulation results, where HSS-GGM achieved reliable false discovery control while retaining competitive power. In the present data analysis, the method similarly identifies a parsimonious set of conditional dependencies while avoiding the markedly denser graphs produced by some methods that do not explicitly control false discoveries.

\subsection{S\&P 100 stock returns}
\label{subsec:finance}

We next illustrate the method on daily stock returns from the S\&P~100 index. We selected \(p=90\) stocks from three sectors, Technology, Healthcare and Financials, with balanced representation across sectors. Daily adjusted closing prices were collected for the period from June~1,~2024 to December~1,~2024, yielding \(n=126\) observations after preprocessing. Log-returns were computed and standardized to have mean zero and variance one. Because stock returns often exhibit heavy tails, we apply the nonparanormal transformation of \citet{Liu2009} before estimating the graphical model. Additional details on data construction, descriptive statistics and assessment of Gaussianity are given in Appendix~\ref{sec:descriptive_finance}.

In this setting, a non-zero entry \(K_{ij}\) in the precision matrix indicates that the returns of stocks \(i\) and \(j\) remain associated after conditioning on all other assets. The resulting graph is therefore a partial correlation network, in which edges represent direct conditional dependencies rather than marginal correlations induced by common market or sectoral factors. Such networks are commonly used in financial econometrics to identify groups of assets with related risk profiles and to study channels of risk transmission \citep{Barigozzi2019,Giudici2020}.

In this application, sector labels are incorporated as group information in the HSS-GGM prior. This reflects the well-known tendency of firms within the same sector to be exposed to similar economic drivers and therefore provides a natural source of prior structure for the estimation of the dependence network.

\begin{figure}[!htbp]
\centering
\includegraphics[width=\textwidth]{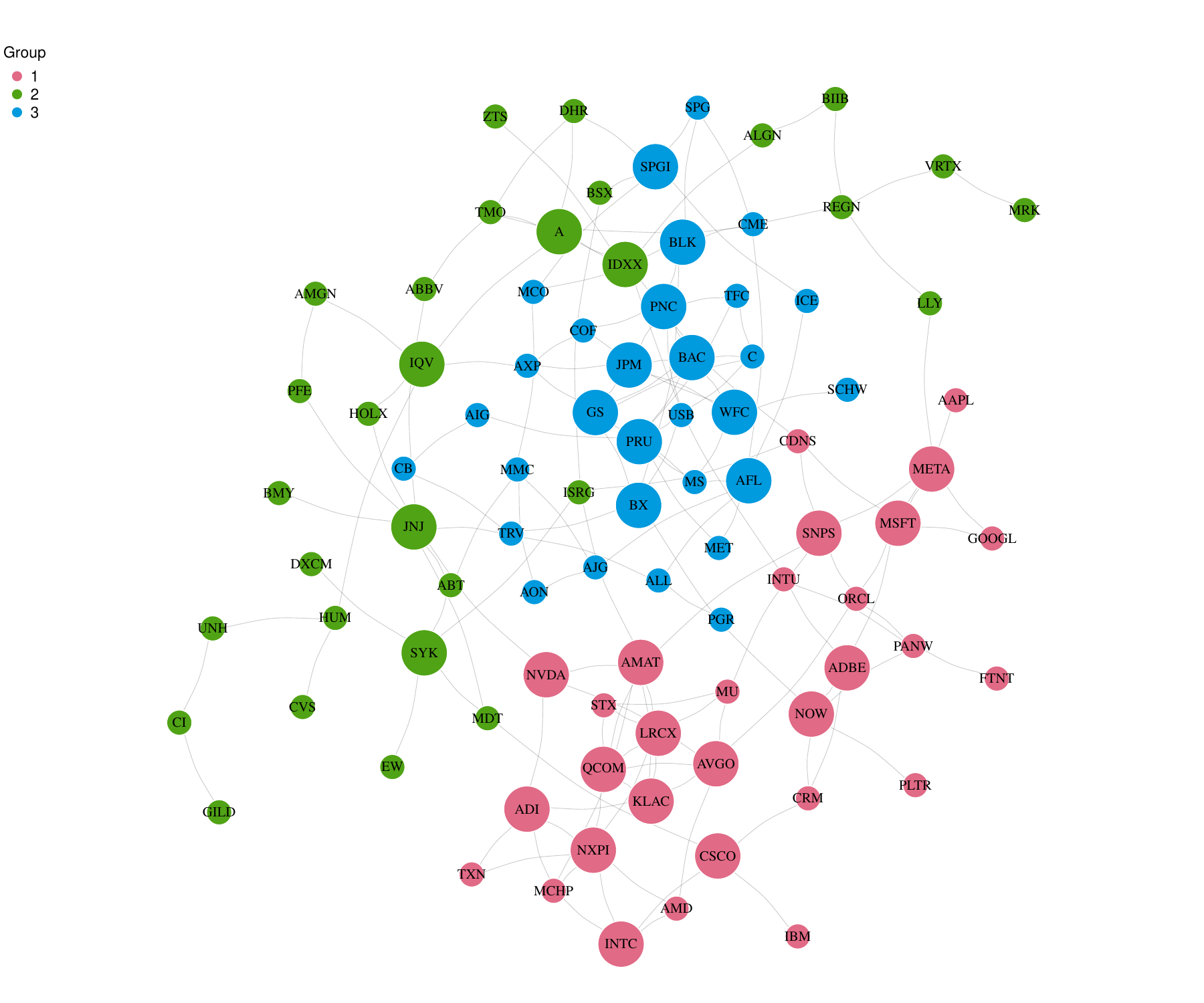}
\caption{Partial correlation network of S\&P~100 stocks inferred by HSS-GGM at nominal level \(\alpha=0.05\). Node colours indicate sector membership: red for Technology, green for Healthcare, and blue for Financials. Edges represent conditional dependencies between stock returns after adjustment for all other assets.}
\label{fig:finance_graph}
\end{figure}

Figure~\ref{fig:finance_graph} displays the network inferred by HSS-GGM at nominal level \(\alpha=0.05\). Because the graph is substantially larger than in the gene expression application, visual comparison with multiple competing methods is less informative, so we focus here on the network produced by HSS-GGM.

The inferred graph shows a clear sectoral organization. Most edges occur within sectors, particularly in Technology and Financials, reflecting shared economic factors within industries. Cross-sector edges are less frequent and tend to involve stocks with relatively high connectivity. Several large-cap firms appear as more central nodes, including MSFT, AVGO and META in Technology, BLK and WFC in Financials, and IDXX and JNJ in Healthcare. These companies are major market participants whose returns often reflect broader sectoral or market-wide information.

Overall, the inferred network is sparse while retaining economically interpretable sector-based structure. This empirical pattern is in line with the simulation results, where HSS-GGM achieved a favourable balance between false discovery control and detection power.

\bibliographystyle{apalike}
\bibliography{BiblioGGM}

\end{document}